\documentclass[fleqn,12pt]{article}
\usepackage{mekit03}
\usepackage{isolatin1}
\usepackage{graphicx}
\usepackage[fleqn]{amsmath}
\usepackage{dcolumn}
\title{Spectral element simulations of buoyancy-driven flow}
\author{Thor Gjesdal%
\thanks{e-mail: thg@ffi.no}, Carl Erik Wasberg, and Øyvind Andreassen}
\address{Norwegian Defence Research Establishment, NO-2027 Kjeller}
\begin{document}
\maketitle
\thispagestyle{empty}%
\begin{abstract}
This paper is divided in two parts.
In the first part, a brief review of a spectral element method for the numerical solution 
of the incompressible Navier-Stokes equations is given.
The method is then extended to compute buoyant flows described by the Boussinesq 
approximation. Free convection in closed two-dimensional cavities are computed and the results
are in very good agreement with the available reference solutions.
\keywords{Spectral element method, incompressible flow, Boussinesq approximation, free convection}
\end{abstract}
\section{Introduction}
At the previous conference~%
\cite{CEWasberg_OAndreassen_BAPReif_2001a}
we presented a method for the numerical solution of the incompressible Navier-Stokes
equations by spectral element methods. In this contribution we will present the current
status of this development effort, in particular new developments to enable computation of 
buoyancy-driven flows described by the Boussinesq approximation.
We will first give a review of the status of the basic Navier-Stokes solver.
Then we describe the Boussinesq model and discuss its numerical implementation.
Simulations of free-convection flows in closed two-dimensional 
cavities show very good agreement with available reference solutions.
\section{Spectral element method}
The Navier-Stokes equations for a constant-density incompressible fluid are
\begin{subequations} \label{NSproblem}
\begin{equation} \label{NSeq}
  \begin{array}{rcll}
    \displaystyle \frac{\partial \mathbf{u}}{\partial t} + \mathbf{u} \cdot \nabla \mathbf{u} & = & \displaystyle - 
    \nabla p + \nu \nabla^2 \mathbf{u}, & \qquad \mbox{in\ } \Omega, \\
    \nabla \cdot \mathbf{u} & = & 0, & \qquad \mbox{in\ } \Omega,
  \end{array} 
\end{equation}
with the following initial- and boundary conditions: 
\begin{equation} \label{NSbc}
  \begin{array}{l}
    \displaystyle \mathbf{u}({\bf x},0) = \mathbf{u}_0({\bf x}), \, {\bf x} \in \Omega, \quad \nabla\cdot \mathbf{u}_0 = 0\\
    \displaystyle \mathbf{u}({\bf x},t) = \mathbf{u}_v({\bf x}, t), \, {\bf x} \in \partial\Omega_v,  \\ 
    \nabla \mathbf{u}({\bf x},t) \cdot {\bf n} = 0, \, {\bf x} \in \partial\Omega_o.
  \end{array}    
\end{equation}
\end{subequations}
The boundary $\partial\Omega$ is divided into two parts
$\partial\Omega_v$ and $\partial\Omega_o$, with Dirichlet velocity
inflow and homogeneous Neumann outflow conditions, respectively, and
${\bf n}$ is the outward pointing normal vector at the boundary.

In~(\ref{NSproblem}), $\Omega \in {\bf R}^d$ is the
computational domain in $d$ spatial dimensions, $\mathbf{u}=\mathbf{u}({\bf x},t)$ is
the $d$-dimensional velocity vector, $p=p({\bf x},t)$ is the kinematic pressure,
and $\nu$ is the kinematic viscosity coefficient. 

To solve~(\ref{NSproblem}) we employ an implicit-explicit time splitting in which
we represent the advective term explicitly, while we treat the diffusive term, 
the pressure term, and the divergence equation implicitly. After semi-discretisation in 
time we can write~(\ref{NSproblem}) in the form
\begin{subequations}
\label{eq:semi-discrete-ns}
\begin{align}
(\alpha I - \nu\nabla^2) \mathbf{u}^{n+1} &= \nabla p + f(\mathbf{u}^{n}, \mathbf{u}^{n-1}, \dots),\\
\nabla\cdot \mathbf{u}^{n+1} &= 0,
\end{align}
\end{subequations}
in which the explicit treatment of the advection term is included in the source term $f$.
In the actual implementation 
we use the BDF2 formula for the transient term,
\[
\frac{\partial u}{\partial t} = 
     \frac{3u^{n+1} -4u^{n}+u^{n-1}}{2\Delta t} + O(\Delta t^2),
\]
which gives $\alpha=3/2\Delta t$ in~(\ref{eq:semi-discrete-ns}),
while we compute the advective contributions according to the 
operator-integration-factor (OIF) method~%
\cite{YMaday_APatera_EMRonquist_1990a}. 

The spatial discretisation is based on a spectral element method~%
\cite{ATPatera_1984a};
the computational domain is sub-divided into non-overlapping quadrilateral
(in 2D) or hexahedral (in 3D) cells or elements. 
Within each element, a weak representation of~(\ref{eq:semi-discrete-ns})
is discretised by a Galerkin method in which we choose the test and trial functions from
 bases of Legendre polynomial spaces
\begin{subequations}
\label{eq:polspaces}
\begin{align}
u^{h}_{i} &\in P_{N}(x) \otimes P_{N}(y) \otimes P_{N}(z), \\
p^{h} &\in P_{N-2}(x) \otimes P_{N-2}(y) \otimes P_{N-2}(z).
\end{align}
\end{subequations}
Note that we employ a lower order basis for the pressure spaces to avoid spurious
pressure modes in the solution. The velocity variables are $C^{1}$-continuous across
element boundaries and are defined in the Legendre-Gauss-Lobatto points for the 
numerical integration, whereas the pressure variable is piecewise discontinuous across
element boundaries and are defined in the interior Legendre-Gauss points.

For the spatial discretization we now introduce the
discrete Helmholtz operator,
\[H = \frac{3}{2\Delta t} B + \nu A, \] 
where $A$ and $B$ are the stiffness- and mass matrices in $d$ spatial dimensions, 
the discrete divergence operator, $D$, and the discrete gradient operator, $G$. 
Appropriate boundary conditions should be included in these discrete operators.
This gives the discrete equations
\begin{subequations}
\label{eq:fully-discrete-ns}
\begin{align} 
  Hu^{n+1} - Gp^{n+1}  &= Bf, \label{eq:fully-discrete-momentum}\\ 
  -Du^{n+1}            &= 0, \label{eq:fully-discrete-continuity}
\end{align}
\end{subequations}
where the change of sign in the pressure gradient term is caused by
an integration by parts in the construction of the weak form of the problem.
This discrete system is solved efficiently by a second order accurate pressure 
correction method that can be written
\begin{subequations}
\label{eq:pressure-correction-method}
\begin{align} 
    Hu^* & =  Bf + Gp^n + r, \\
    DQG(p^{n+1}-p^n) & = - Du^*  \\  
    u^{n+1} & =  u^* + QG(p^{n+1} - p^n),
\end{align}
\end{subequations}
where $u^*$ is an auxiliary velocity field that does not satisfy the 
continuity equation~(\ref{eq:fully-discrete-continuity}).

The discrete Helmholtz operator is symmetric and diagonally dominant, 
since the mass matrix of the Legendre discretisation is diagonal, 
and can be efficiently solved by the conjugate gradient method with a
diagonal (Jacobi) preconditioner.
The pressure operator $DQG$ is easily computed; it is also symmetric, but
ill-conditioned. The pressure system is solved by the
preconditioned conjugate gradient method, with a multilevel overlapping Schwarz
preconditioner~\cite{PFFischer_NIMiller_FMTufo_2000a}.

Earlier~\cite{CEWasberg_OAndreassen_BAPReif_2001a} we presented a validation
of the method for two-dimensional examples. Since then we have extended the 
method to compute the full three-dimensional Navier-Stokes equations. 
At present, turbulence simulations of a fully developed channel flow at 
$\mathrm{Re}_{\tau}=180$ is in progress. Results of these simulations will be 
presented elsewhere.

\begin{figure}
\begin{center}
\includegraphics{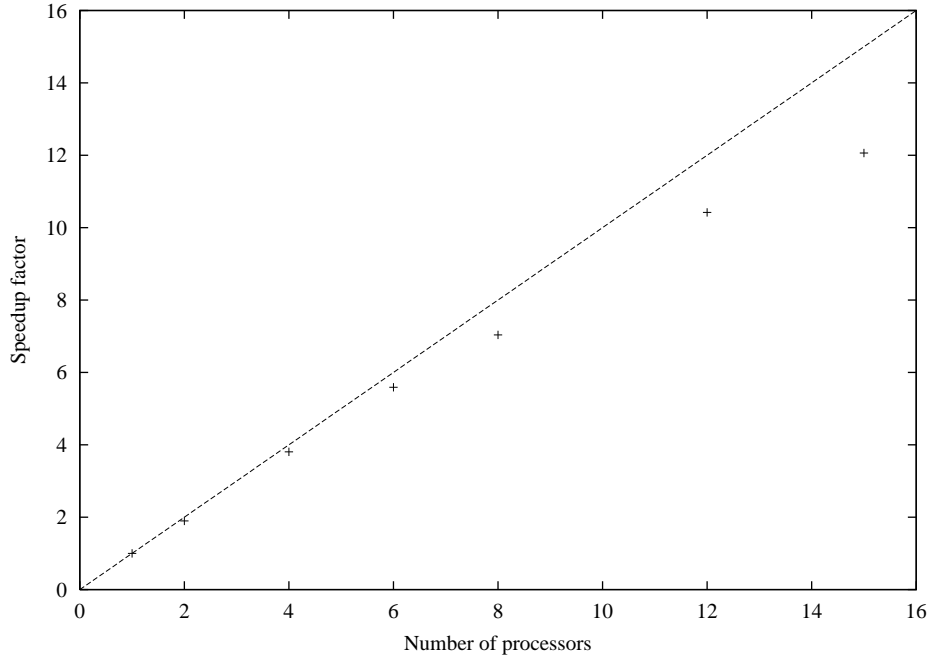}
\end{center}
\caption{
Parallel speed-up for simulation of a three-dimensional fully developed 
turbulent channel flow.
\label{fig:speedup}
}
\end{figure}

The method has good data locality and can be efficiently run on parallel computers.
We have parallelized the code by message passing (MPI) which enables execution
on both distributed-memory cluster and shared-memory architectures. 
To demonstrate the parallel performance, we show in Fig.~%
\ref{fig:speedup}
the speed-up factors for a Direct Numerical Simulation of three-dimensional fully
developed turbulent channel flow using approximately 250000 grid points. 
The computations were performed on a 16 processor SGI system.

\section{Solution method for buoyant flow}
The equations describing the dynamics of incompressible, viscous, buoyant flows under the Boussinesq
approximation are
\begin{subequations}
\label{eq:boussinesq}
\begin{eqnarray}
\nabla\cdot\mathbf{u} &=& 0, \\
\frac{\partial \mathbf{u}}{\partial t} + \mathbf{u}\cdot\nabla{\mathbf{u}} &=& 
   -\nabla{p} + \nu\nabla^2 \mathbf{u} + \beta\left(T-T_\mathrm{ref}\right)\mathbf{g}, \\
\frac{\partial T}{\partial t} + \mathbf{u}\cdot\nabla{T} &=& \alpha\nabla^2 T, 
\end{eqnarray}
\end{subequations}
where $T$ represents the temperature, $\alpha$ the thermal diffusivity, and $\beta$
the coefficient of thermal expansion.
The Boussinesq approximation is valid provided that the density variations, 
$\rho(T)$,  are small; in practice this means that that only small temperature
deviations from the mean temperature are admitted.

The relevant non-dimensional groups to characterize the flow are:
\begin{itemize}
\item{The Prandtl number  $\mathrm{Pr}=\nu/\alpha$,}
\item{the Reynolds number $\mathrm{Re}=UL/\nu$, and}
\item{the Rayleigh number $\mathrm{Ra}=g\beta\Delta TL^3/\nu\alpha$.}
\end{itemize}
Note that the Reynolds number is only relevant for problems with an imposed velocity scale.
The free convection problems we consider below are completely determined by the Prandtl and
Rayleigh numbers.

In this section we will describe the solution method for the Boussinesq problem. Note that the buoyancy 
effect is accounted for in \eqref{eq:boussinesq} through the solution of an additional scalar 
advection-diffusion equation and an extra source term in the momentum equations.

The key to efficient and accurate solution of the Boussinesq/Navier-Stokes system is to use
an implicit-explicit splitting of diffusive and advective terms.
In particular, if the advection/diffusion equations are solved by an implicit/explicit procedure, 
the temperature equation can be decoupled from the remaining Navier-Stokes equations, and the 
buoyancy source term can be calculated first and fed directly to the Navier-Stokes solver.
For illustrative purposes we will discuss the solution procedure in term of an implicit-explicit
first order Euler time discretization. Note however that higher order methods and operator splitting
are used in the actual implementation as discussed for the basic Navier-Stokes solver above.
A first order semi-discrete solution of the Boussinesq system can be written
\begin{subequations}
\label{eq:1st-order-semi-discretisation}
\begin{eqnarray}
\frac{ T^{n+1}-T^{n} }{ \Delta t } +\left(\mathbf{u}\cdot\nabla T  \right)^{n} &=& 
                                                                   (\kappa\nabla^2 T)^{n+1}, \\
\frac{ \mathbf{u}^{n+1}-\mathbf{u}^{n} }{ \Delta t } +
   \left(\mathbf{u}\cdot\nabla \mathbf{u}  \right)^{n} &=& -\nabla p^{n+1} +
   (\kappa\nabla^2 \mathbf{u})^{n+1}  
   +\alpha\left(T^{n+1}-T_\mathrm{ref}\right)\mathbf{g},\\
\left(\nabla\cdot\mathbf{u}\right)^{n+1} &=& 0,
\end{eqnarray}
\end{subequations}
where we have changed the ordering of the equations to emphasize that the temperature at the new
time level, $T^{n+1}$, can be obtained from old velocity data since the advection term is treated
explicitly. A possible solution algorithm is then
self-evident:
\begin{enumerate}
\item{Solve the advection-diffusion equation to obtain $T^{n+1}$.}
\item{Calculate the buoyancy source term.}
\item{Calculate the explicit contributions to the momentum equations.}
\item{Solve the remaining Stokes problem with the pressure 
      correction method to obtain $\mathbf{u}^{n+1}$ and $p^{n+1}$.}
\end{enumerate}

The actual implementation is based on higher-order methods and
the operator integration factor splitting method described above for the 
advection/diffusion equations (both for temperature and momentum). The method uses second order accurate
integrators; for the advection terms we use an adaptive Runge-Kutta method, while the implicit parts are 
solved by the second order implicit Euler scheme (BDF2).
\section{Simulations of free convection cavities}
We have performed simulation of the free convection in two-dimensional square and rectangular cavities.
Cavity flows are often used as test cases for code validation, because they are simple to set up
and reliable reference solutions are readily available. Furthermore, thermal cavity flows 
display a plethora of interesting fluid dynamic phenomena, and they are important prototype
flows for a wide range of practical technological problems, including ventilation,
crystal growth in liquids, nuclear reactor safety, and the design of high-powered laser systems.
\subsection{Differentially heated square cavity}
The steady-state differentially heated square cavity flow was the subject of one of the first benchmark comparison
exercises reported in~\cite{GdeVahlDavis_IPJones_1983a}. The benchmark results produced in that exercise
are given in~\cite{GdeVahlDavis_1983a}. The results of de Vahl Davis were produced, for Rayleigh
numbers in the range $10^3$--$10^6$, using a stream-function/vorticity formulation discretised by a 
second-order finite difference method on a regular mesh. 
Later, more accurate results obtained by a second order finite volume method on higher resolution 
non-uniform grids were presented in~\cite{hortmann-etal:multigrid-benchmark-90}. 

The problem comprises a square box of side length $L_x=L_y=L$ filled with a Boussinesq fluid characterized
by a Prandtl number, $\mathrm{Pr}=0.71$. The vertical walls are kept at a constant temperature,
$T_{\mathrm{hot}}$ and $T_{\mathrm{cold}}$, respectively while the horizontal lid and bottom are
insulated with zero heat flux. The direction of gravity is downwards, i.e. in the negative
$y$-direction. 

\begin{figure}
\includegraphics[scale=0.35]{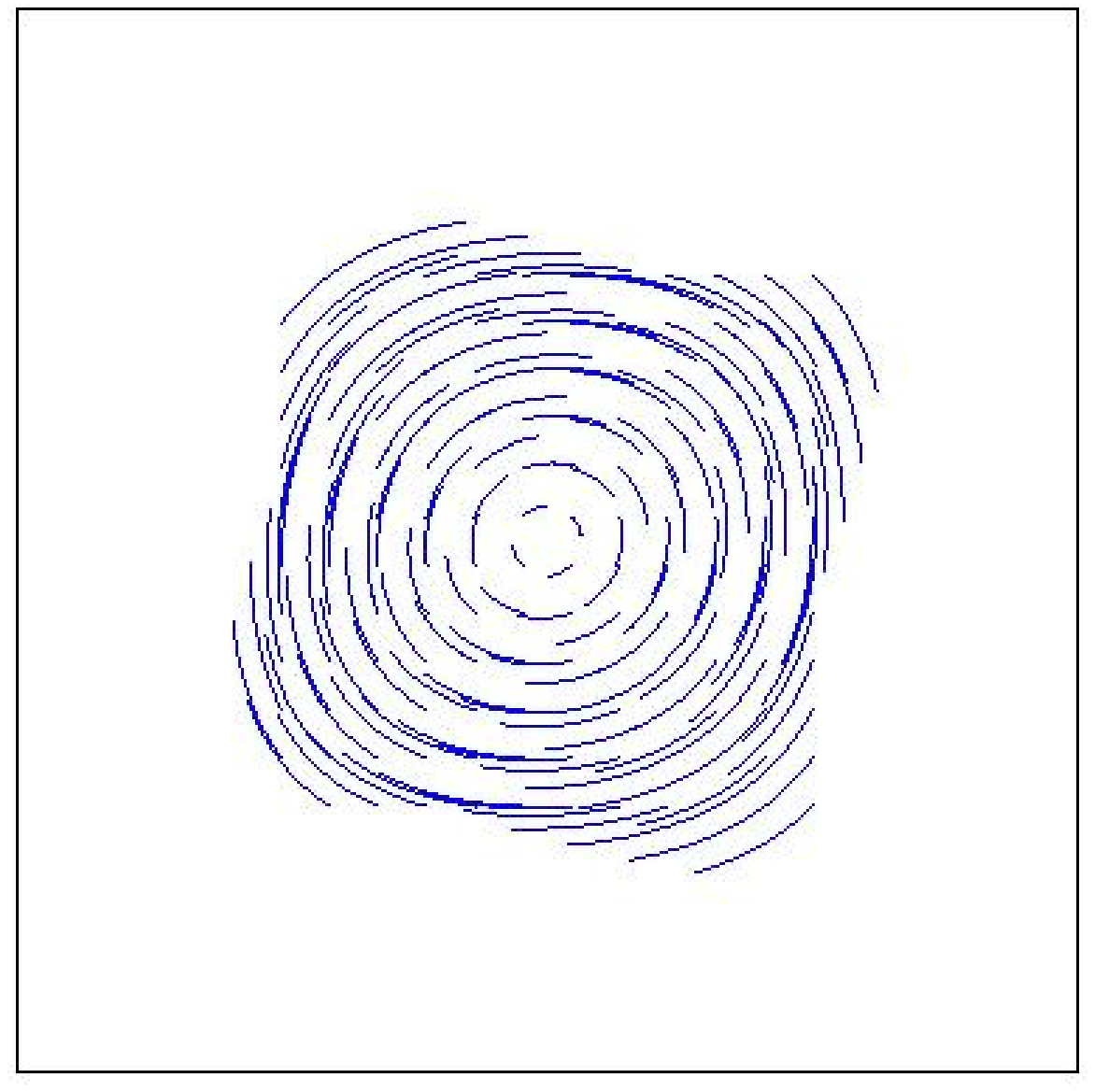}
\includegraphics[scale=0.35]{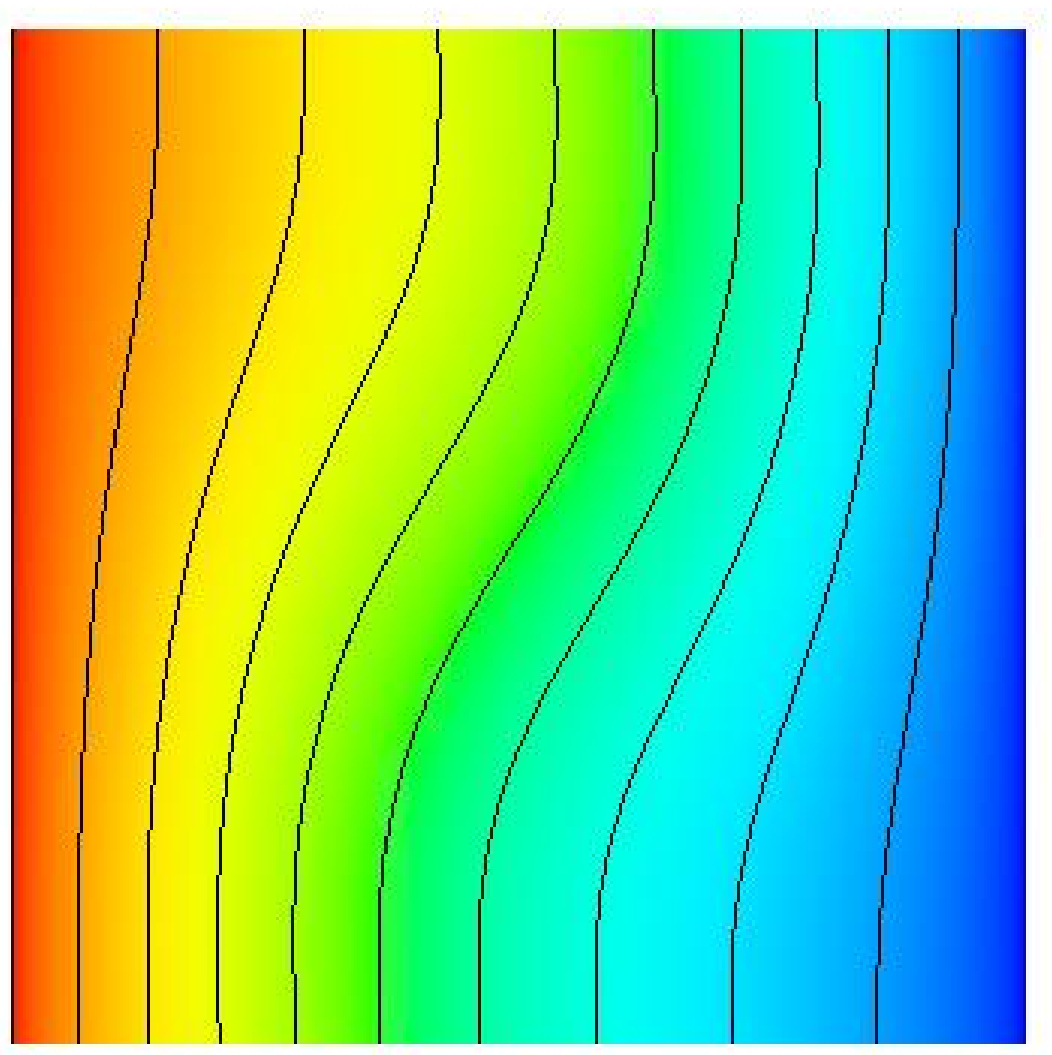}
\caption{Streamlines and temperature distribution for the side-heated buoyant cavity flow at $Ra=10^{3}$.
         \label{fig:side-cavity-1e3}
}
\end{figure}

\begin{figure}
\includegraphics[scale=0.35]{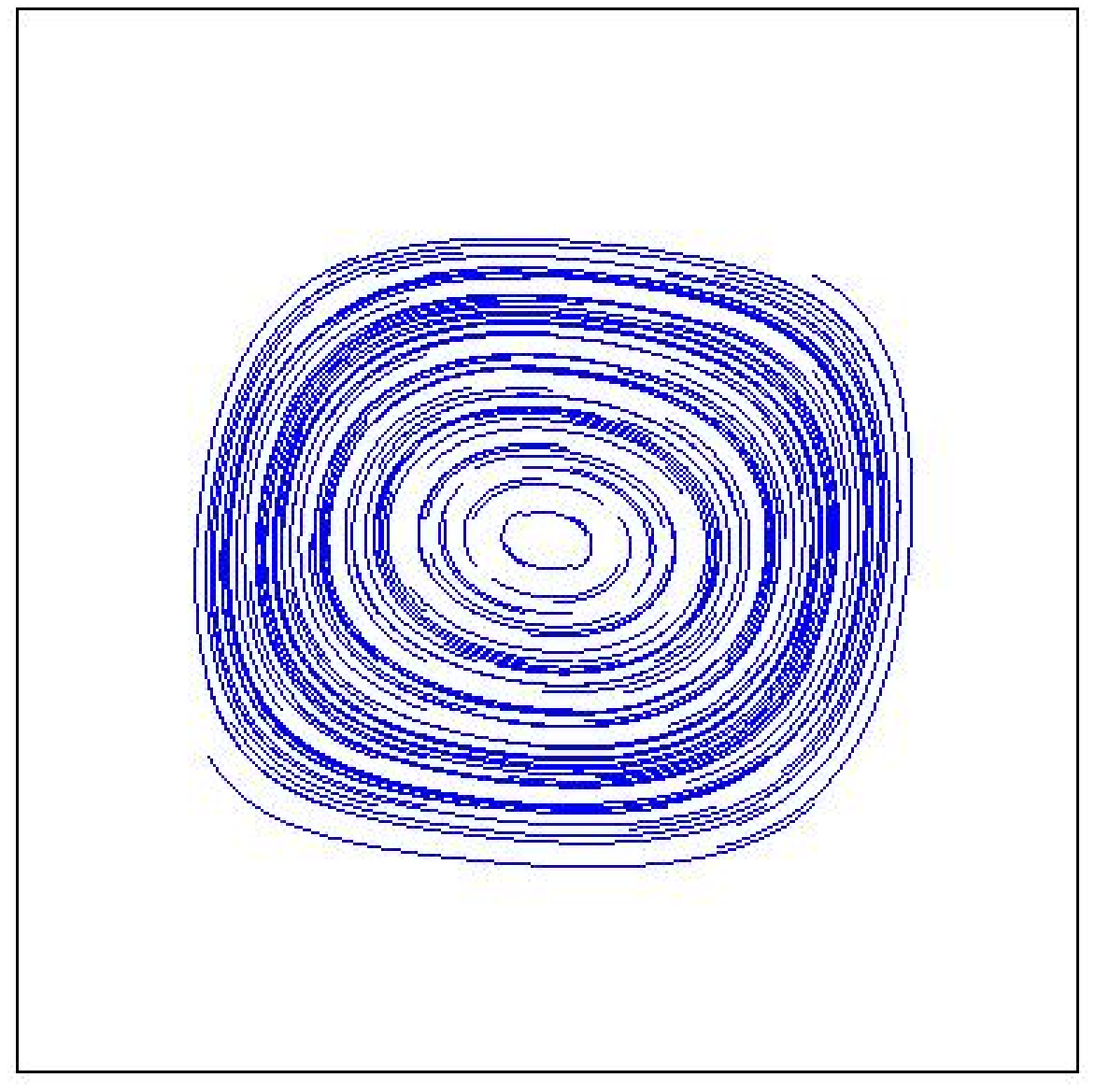}
\includegraphics[scale=0.35]{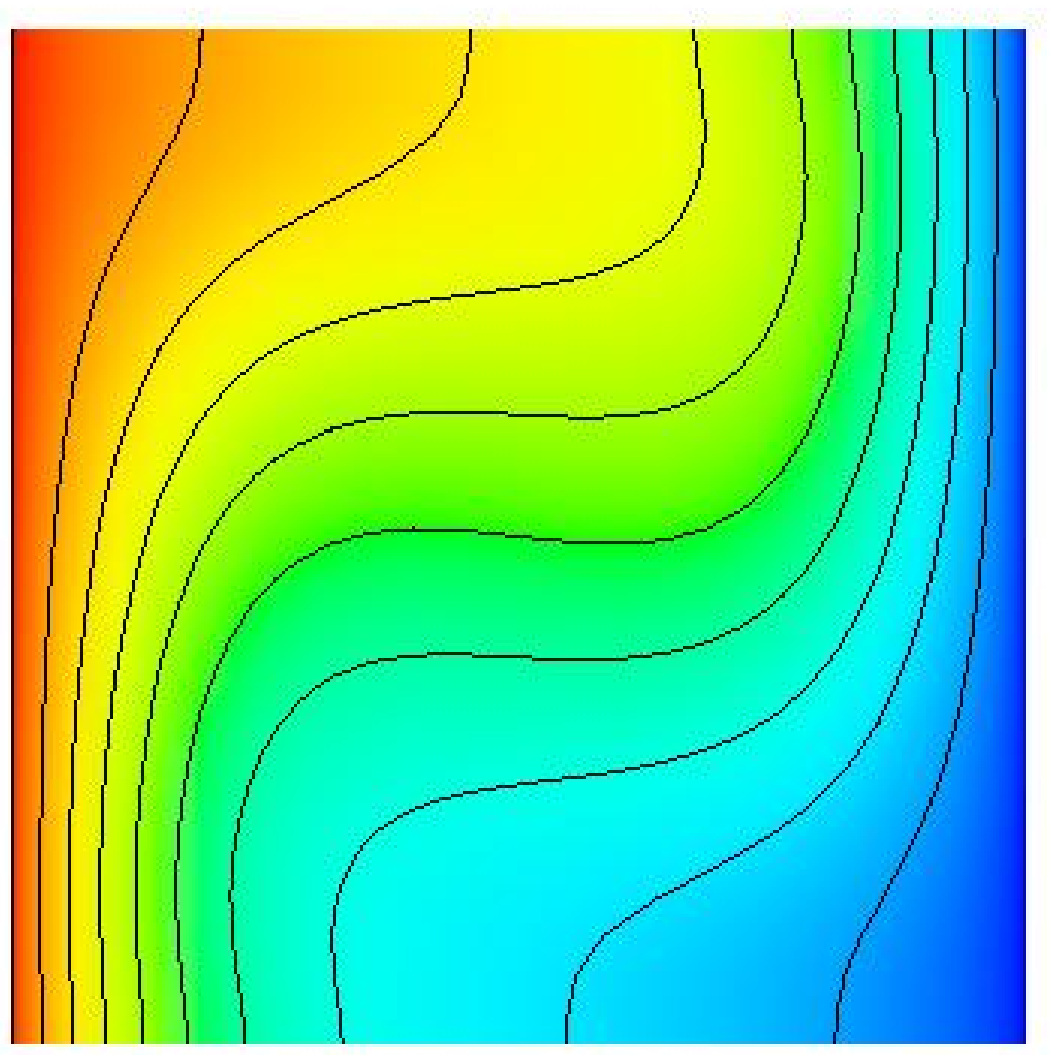}
\caption{Streamlines and temperature distribution for the side-heated buoyant cavity flow at $Ra=10^{4}$.
         \label{fig:side-cavity-1e4}
}
\end{figure}

\begin{figure}
\includegraphics[scale=0.35]{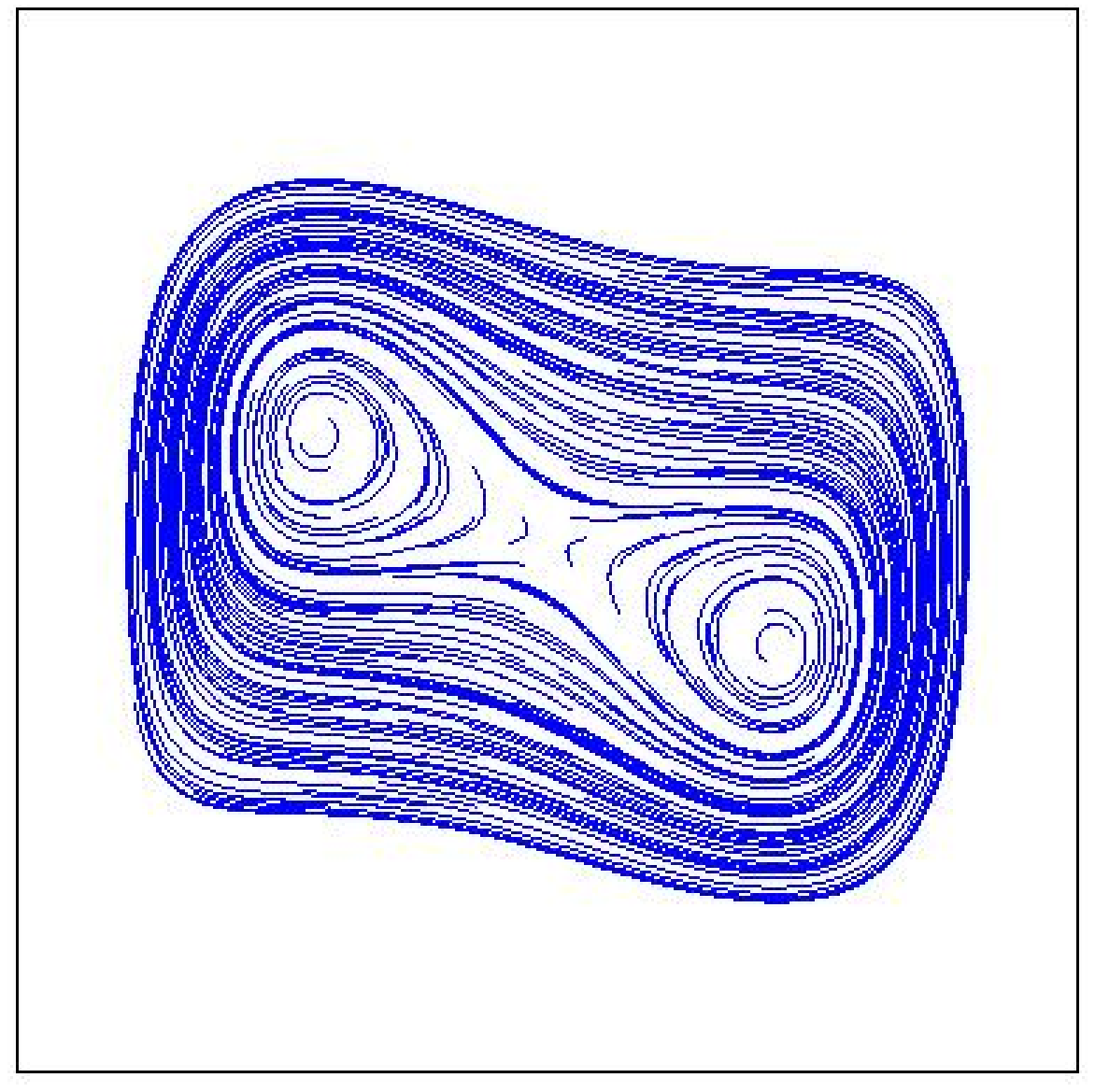}
\includegraphics[scale=0.35]{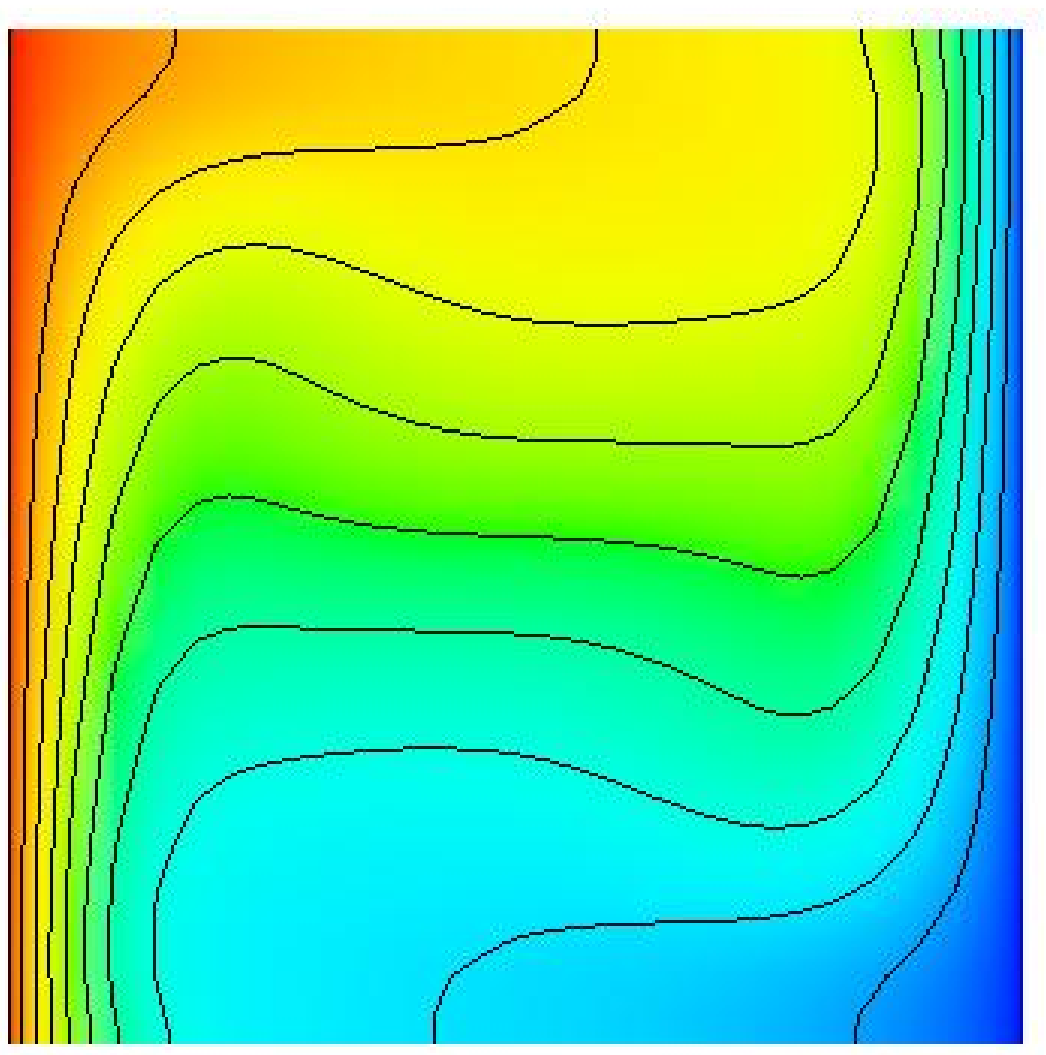}
\caption{Streamlines and temperature distribution for the side-heated buoyant cavity flow at $Ra=10^{5}$.
         \label{fig:side-cavity-1e5}
}
\end{figure}

\begin{figure}
\includegraphics[scale=0.35]{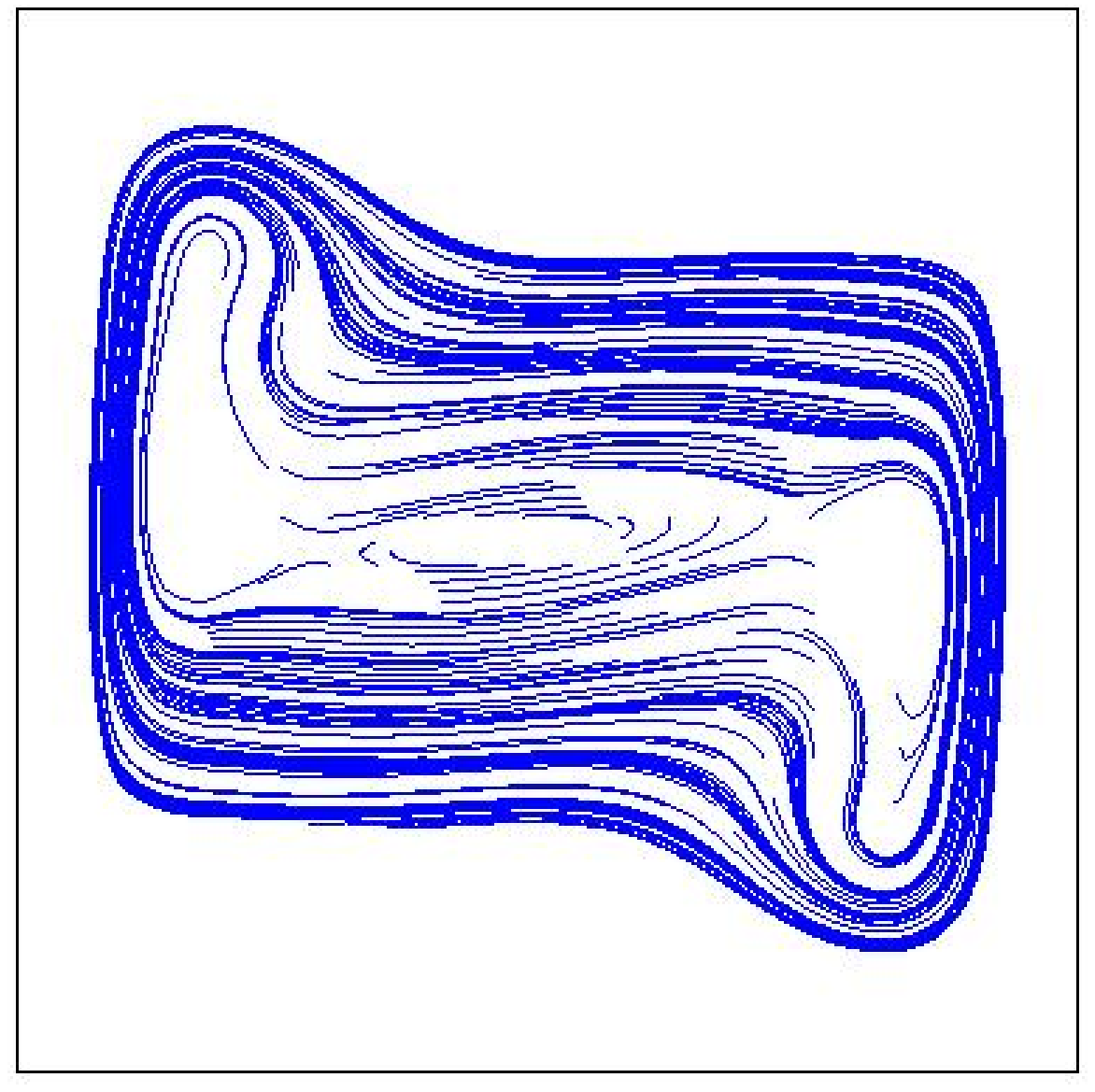}
\includegraphics[scale=0.35]{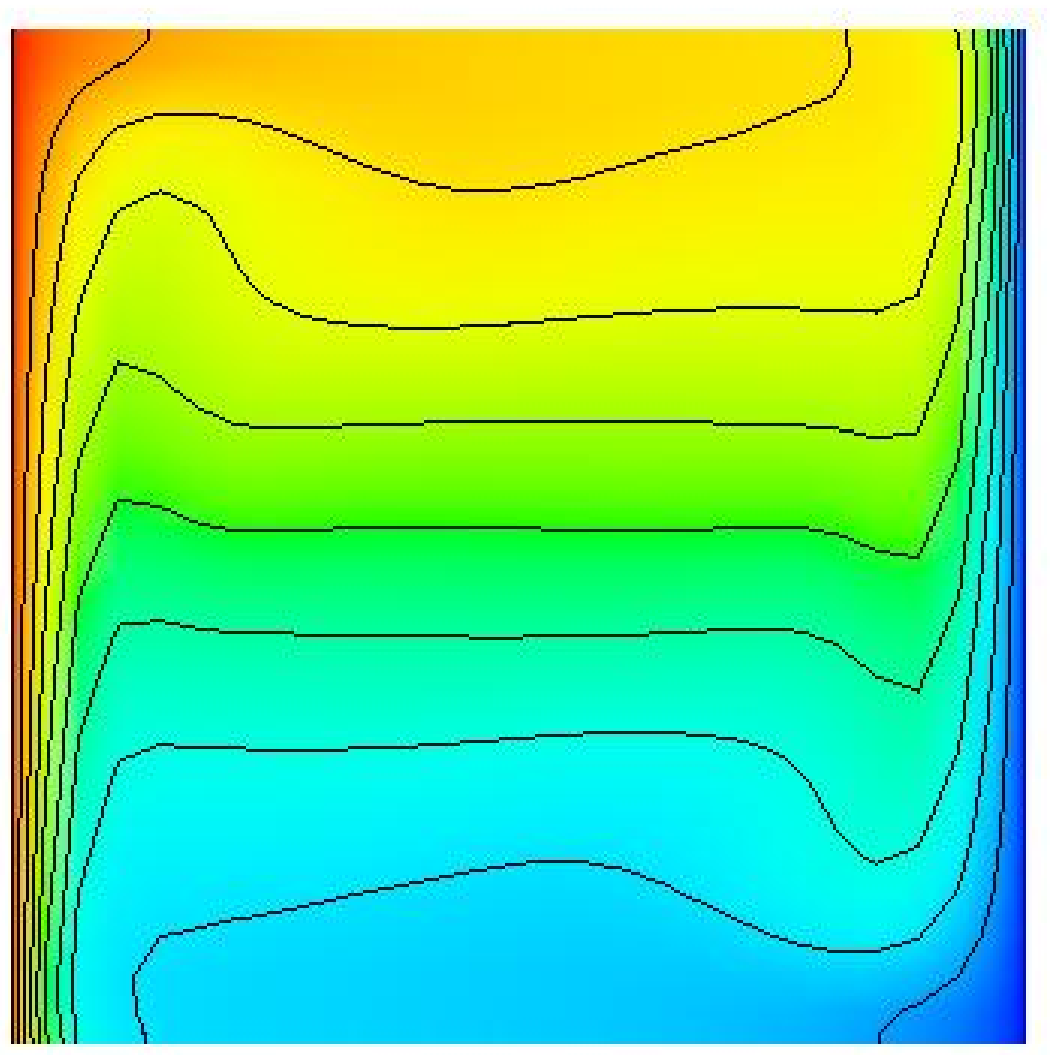}
\caption{Streamlines and temperature distribution for the side-heated buoyant cavity flow at $Ra=10^{6}$.
         \label{fig:side-cavity-1e6}
}
\end{figure}

We performed calculations for $10^3\leq\mathrm{Ra}\leq10^6$, and at these Rayleigh numbers the flow is stationary.
We show the computed flow field and temperature distributions in Figs.~%
\ref{fig:side-cavity-1e3}--\ref{fig:side-cavity-1e6}.

The most important diagnostic connected to the free convection cavity flow is the average
Nusselt number, which expresses the non-dimensional heat flux across the cavity.
The Nusselt number is usually calculated at a vertical line, typically the hot and the cold wall. 
For consistency with the weak Galerkin formulation, we have however chosen to compute a 
global Nusselt number given by
\begin{subequations}
\label{eq:nusselt-number}
\begin{equation}
   \mathrm{Nu} = \frac{Q}{Q_0},
\end{equation}
where $Q$ is the calculated global heat flux through the cavity
\begin{equation}
Q = {\int_0^{L_{x}}{ \int_0^{L_{y}}{  uT - \alpha\frac{\partial T}{\partial x}dxdy}}},
\end{equation} 
and the reference value, $Q_0$, is the corresponding heat flux if the heat transfer were by
pure conduction
\begin{equation}
  Q_0 = {L_{x} L_{y}}\frac{\alpha\Delta T}{L_x} = L_{y}\alpha\Delta T.
\end{equation}
\end{subequations}
We have confirmed that the computed values of the average global Nusselt number does indeed
agree with the average wall Nusselt numbers. 

\begin{figure}
\includegraphics[]{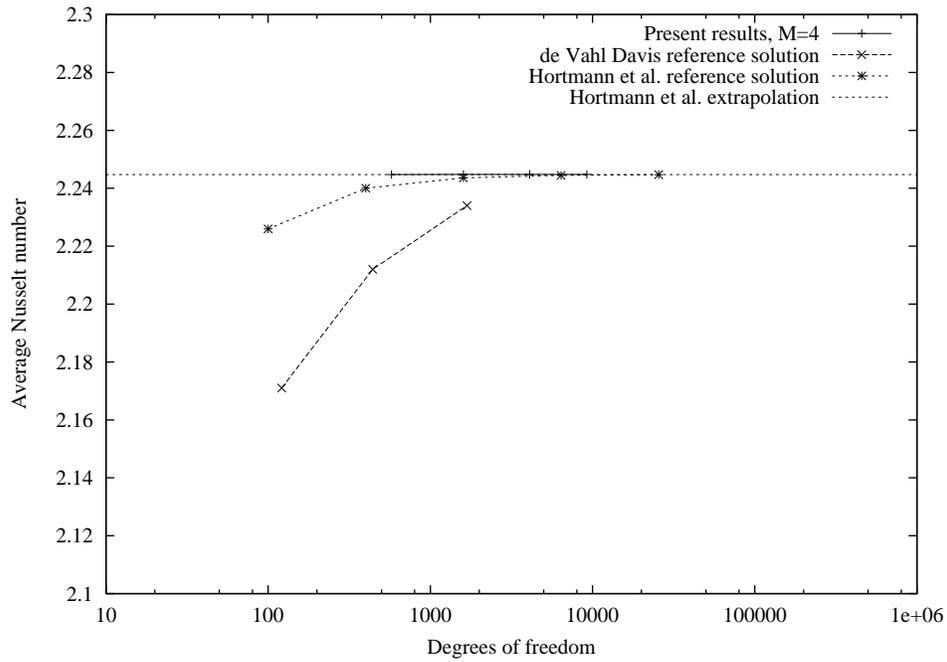}
\caption{Grid convergence of the average Nusselt number for the 
         differentially heated buoyant cavity flow at $\mathrm{Ra}=10^{4}$.
         \label{fig:nusselt-number-sqaure-cavity-1e4}
}
\end{figure}

\begin{figure}
\includegraphics[]{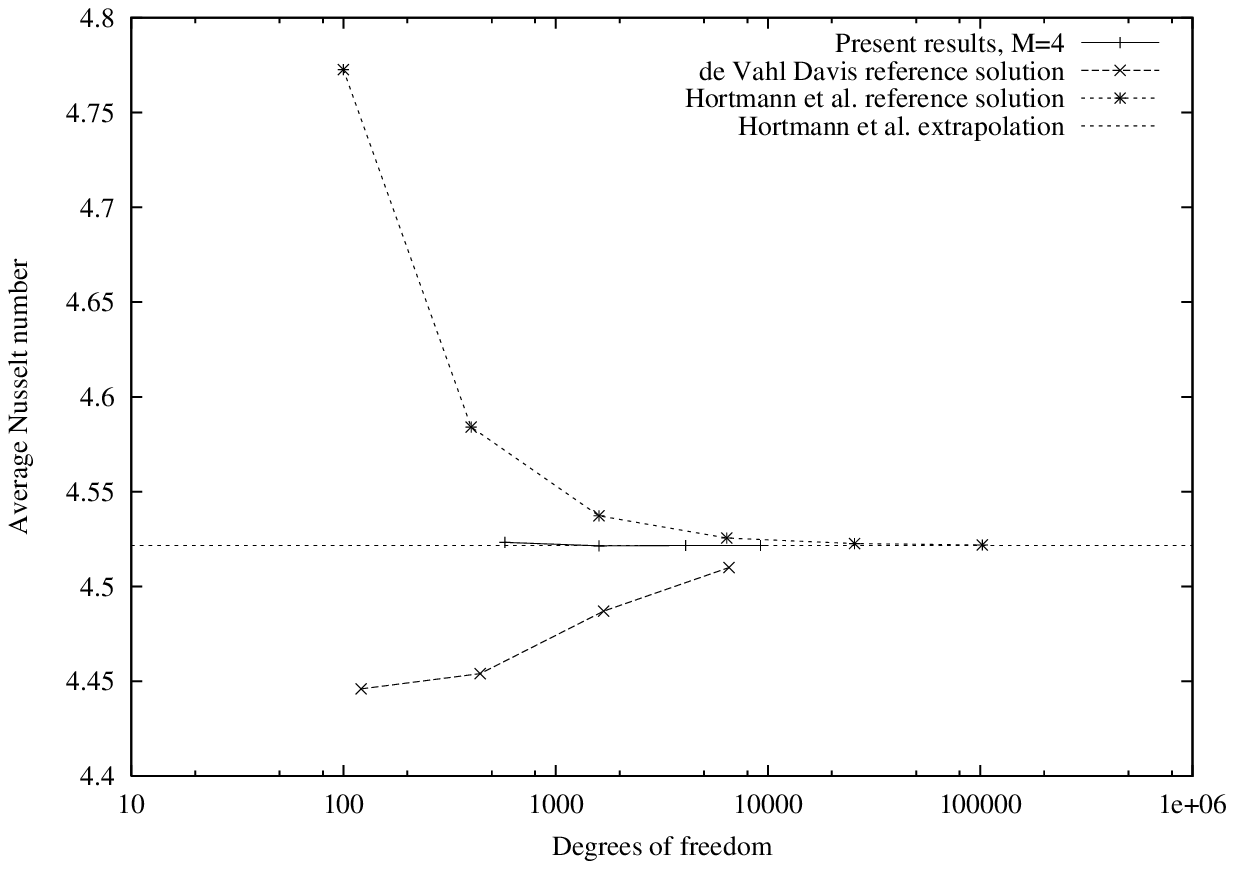}
\caption{Grid convergence of the average Nusselt number for the 
         differentially heated buoyant cavity flow at $\mathrm{Ra}=10^{5}$.
         \label{fig:nusselt-number-sqaure-cavity-1e5}
}
\end{figure}

\begin{figure}
\includegraphics[]{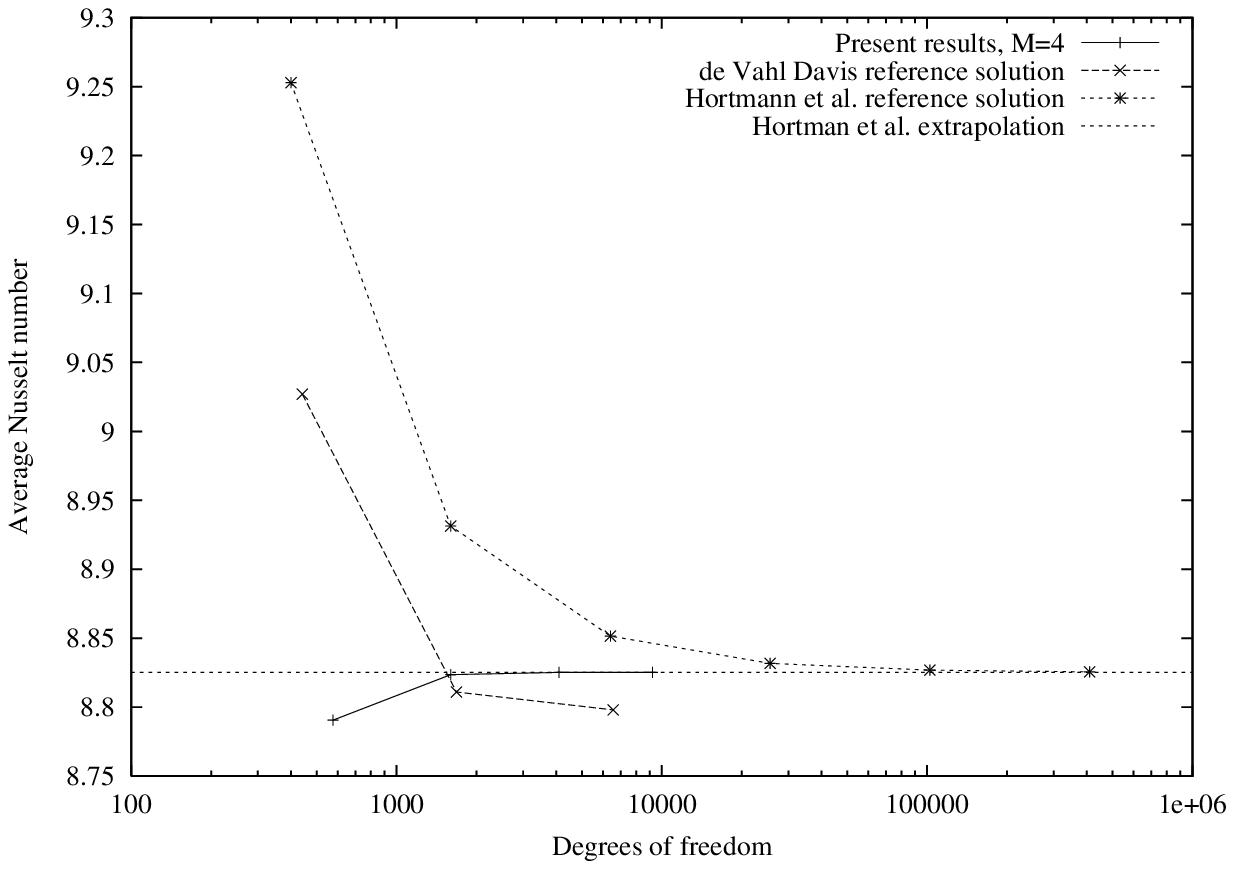}
\caption{Grid convergence of the average Nusselt number for the 
         differentially heated buoyant cavity flow at $\mathrm{Ra}=10^{6}$.
         \label{fig:nusselt-number-sqaure-cavity-1e6}
}
\end{figure}

We performed simulations using $M=4\times 4$ elements varying the resolution in each
element from $N=6\times 6$ to $N=24\times 24$.
In Figs.~%
\ref{fig:nusselt-number-sqaure-cavity-1e4}--\ref{fig:nusselt-number-sqaure-cavity-1e6}
we show the grid convergence of the computed Nusselt numbers compared to the previously
reported benchmark results~%
\cite{GdeVahlDavis_1983a} and \cite{hortmann-etal:multigrid-benchmark-90}.
Note the excellent agreement with the reference data; even the coarsest resolution
(i.e. $24\times 24$) produces solutions that are essentially converged except at the highest
Rayleigh number.
In Table~\ref{table:square-cavity-nusselt-numbers} we compare the Nusselt numbers obtained
at the finest grid with the `grid-independent' values from the reference
solutions obtained by Richardson extrapolation.
\begin{table}
\caption{Computed Nusselt numbers for the square cavity compared to the 
         extrapolated results of the reference solutions of de Vahl Davis (1983)
         and Hortmann \emph{et al.} (1990)\label{table:square-cavity-nusselt-numbers}
}
\begin{center}
\begin{tabular}{lD{.}{.}{3}D{.}{.}{3}D{.}{.}{3}}
Rayleigh no:               & \multicolumn{1}{c}{$10^4$} & \multicolumn{1}{c}{$10^5$} & \multicolumn{1}{c}{$10^6$} \\ \hline
Present results: & 2.245                 &  4.522                  &  8.825                  \\
de Vahl Davis  : & 2.243                 &  4.519                  &  8.800                  \\
Hortmann \emph{et al.}: 
                 & 2.245                 &  4.522                  &  8.825                  \\ \hline
\end{tabular}
\end{center}
\end{table}
\subsection{Bottom-heated square cavity}
\begin{figure}
\includegraphics[]{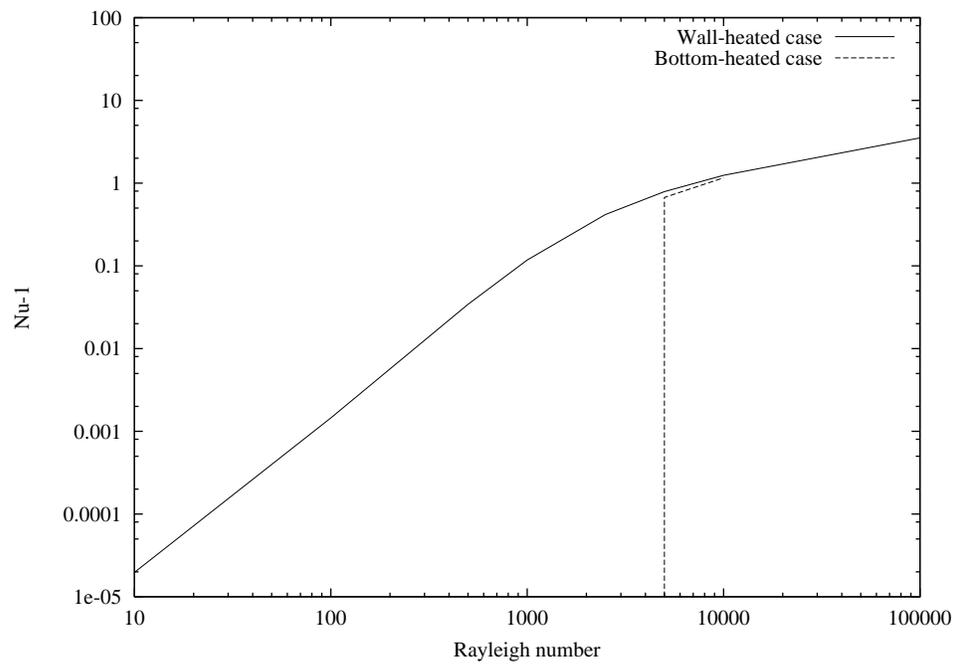}
\caption{Deviatory Nusselt number for the square cavity heated from below.
         \label{fig:bottom-square-cavity-deviatory-nusselt}
}
\end{figure}

\begin{figure}
\begin{center}
\includegraphics[scale=0.35,angle=90]{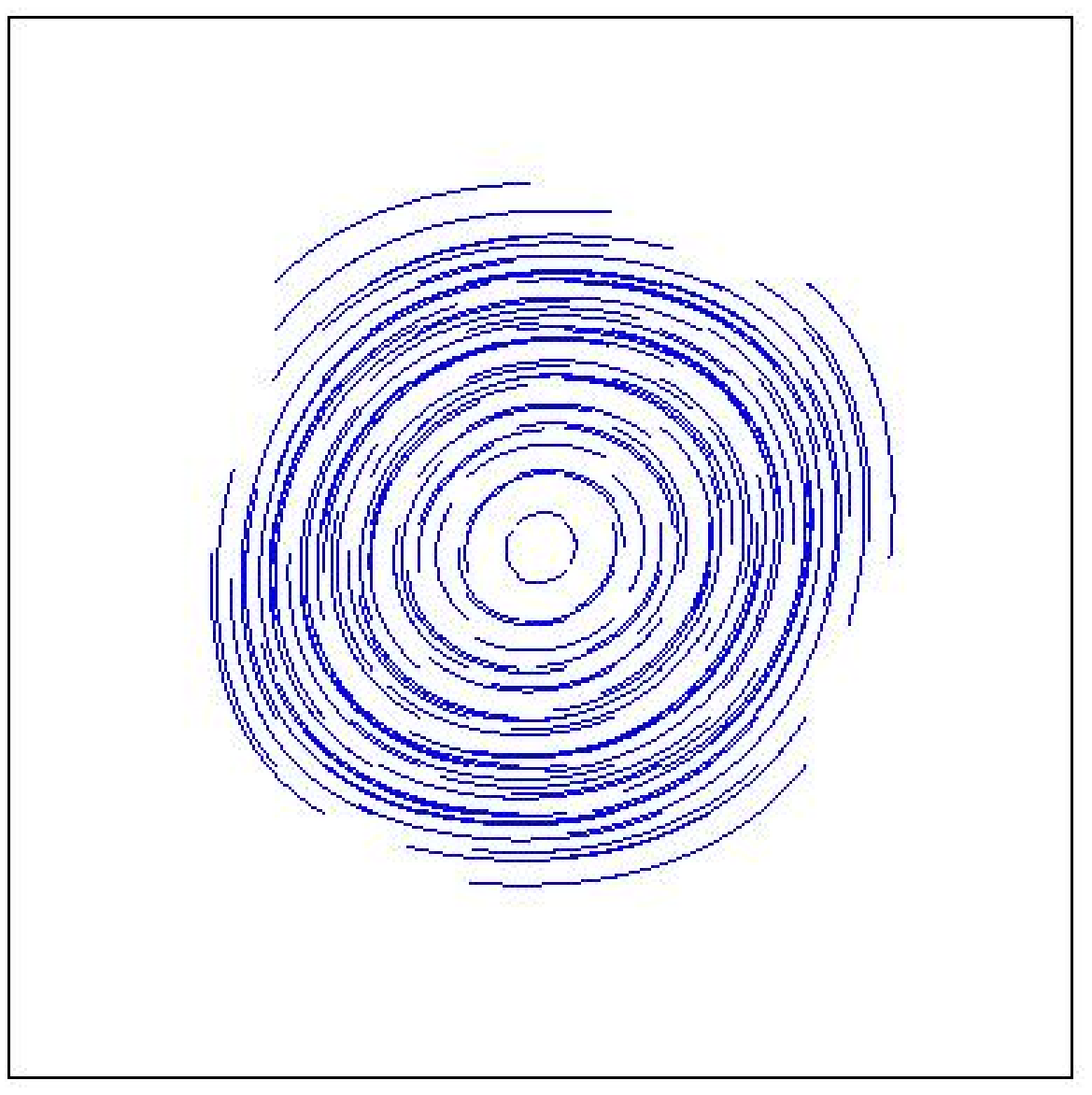}
\includegraphics[scale=0.35,angle=90]{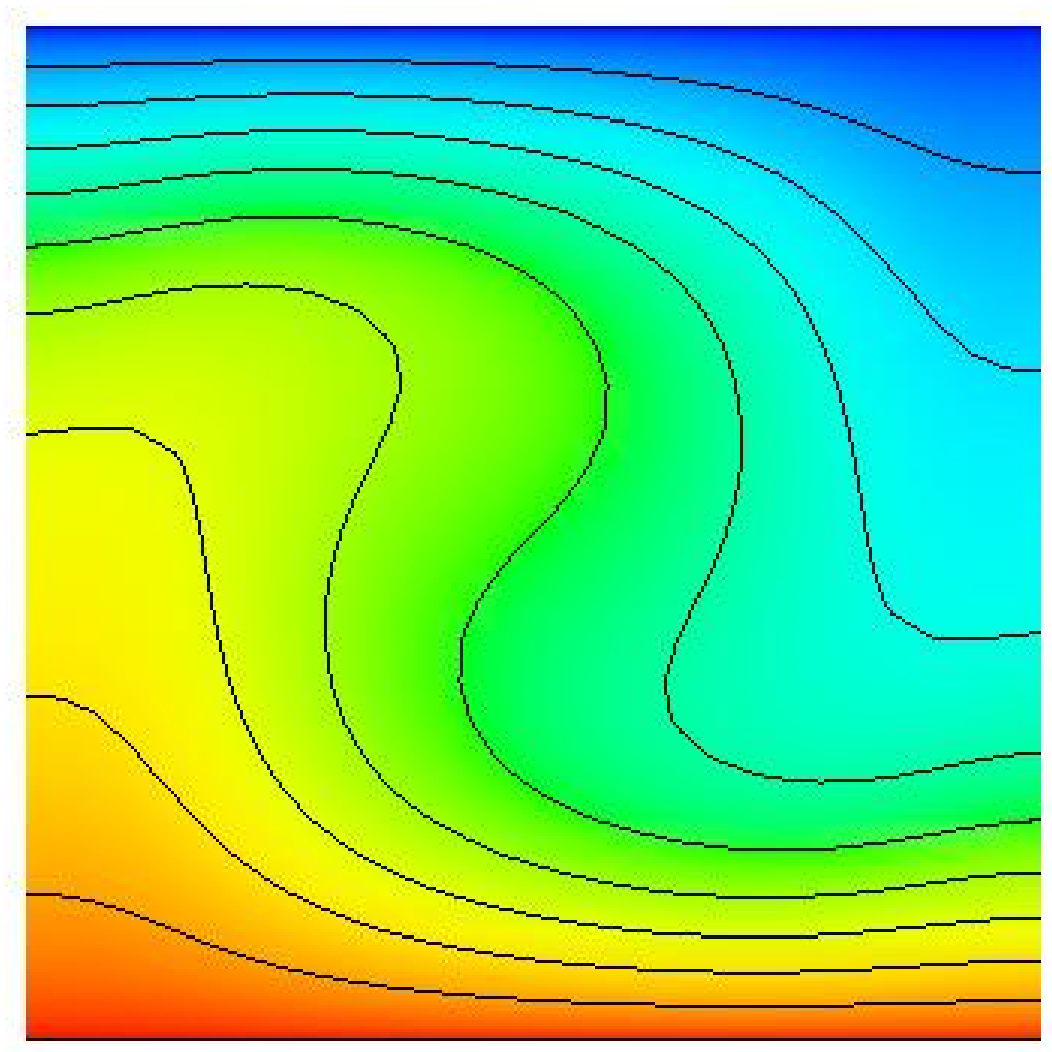}
\end{center}
\caption{Streamlines and temperature distribution for the bottom-heated buoyant cavity flow at $Ra=10^{4}$.
         \label{fig:bottom-cavity-1e4}
}
\end{figure}

It is also interesting to consider the case in which the cavity 
is heated from below instead of from the vertical walls as in the above examples.
When the heated walls are aligned with the direction of gravity the circulation, and
hence convection, is set up at small Rayleigh numbers. Although the bottom-heated case
corresponds to a genuinely unstable situation; gravity will act against the instability
caused by the temperature difference and produce a regime of pure conduction at low 
$\mathrm{Ra}$. 
To illustrate this we show the effect of conduction expressed by the deviatory Nusselt 
number, i.e.  $\mathrm{Nu} - 1$, for the two cases in Fig.~\ref{fig:bottom-square-cavity-deviatory-nusselt}. 
Note that whereas there is a smooth transition into the convection regime in the
wall-heated case, the bottom-heated case shows a sharp transition point below which
heat transfer is purely by conduction. Above the critical point the difference between the
two cases, both with respect to the heat transfer and to the flow field, is small as we can 
see in Fig.~\ref{fig:bottom-cavity-1e4}.
\subsection{Simulation of a tall cavity}
Christon \emph{et al.}~\shortcite{MAChriston_PMGresho_SBSutton_2002a} summarises the results of a 
workshop discussing the free convection in a tall cavity with aspect ratio 8:1.
The comparison was performed for a Rayleigh number $\mathrm{Ra}=3.4\times 10^{5}$,
which is slightly above the transition point from steady-state to time-dependent flow
at $\mathrm{Ra}\approx 3.1\times 10^{5}$. 
A total of 31 solutions were submitted to the workshop, of these a pseudo-spectral
solution using $48\times 180$ modes~%
\cite{SXin_PLeQuere_2002a}
was selected as the reference solution.

\begin{figure}
\includegraphics[]{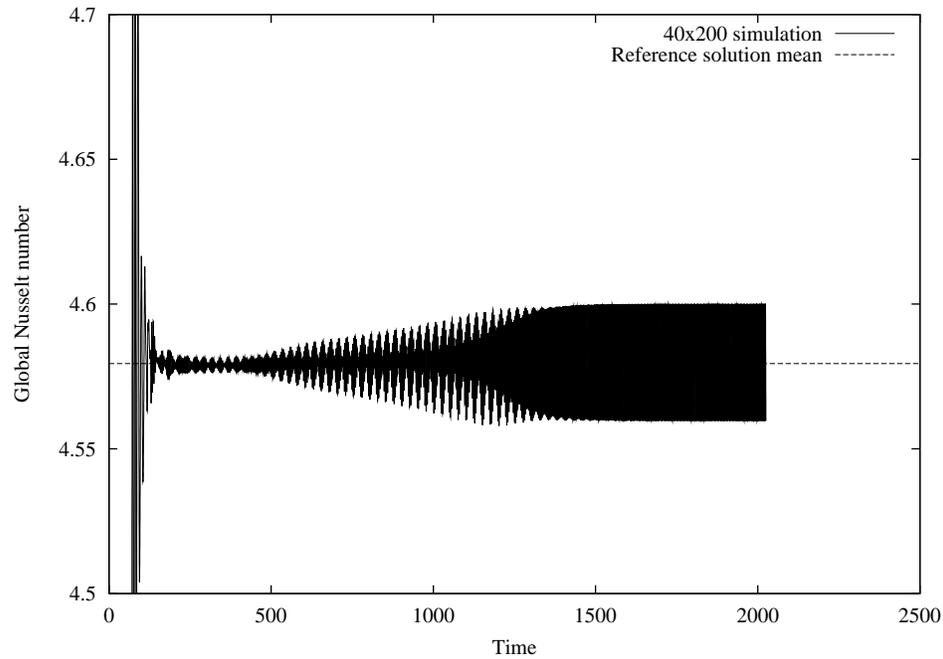}
\caption{Time history for the Nusselt number in the differentially heated tall cavity at 
         $\mathrm{Ra}=3.4\times 10^{5}$.
         \label{fig:nusselt-number-tall-cavity-time-history}
}
\end{figure}

We have computed the solution to this case with roughly the same resolution as in 
the steady-state computations of the square cavity reported above. We show the time
history of the global Nusselt number in Fig.~\ref{fig:nusselt-number-tall-cavity-time-history},
and note that the flow reaches a statistically steady state after approximately
1500 non-dimensional time units
\[
\tau_0 = \sqrt{ \frac{\mathrm{Pr}}{\nu^2 \mathrm{Ra}} }.
\]
Note that there appears to be good agreement with the reference solution mean.
This is confirmed in Table~\ref{table:tall-cavity-metrics} in which we give
time averages of the computed Nusselt number, the velocity metric
\[
U = \sqrt{\frac{1}{2 L_x L_y} {\int_0^{L_{x}}{ \int_0^{L_{y}}{  \mathbf{u}\cdot\mathbf{u} dxdy}}} },
\]
the vorticity metric
\[
\omega = \sqrt{\frac{1}{2 L_x L_y} {\int_0^{L_{x}}{ \int_0^{L_{y}}{  ( v_x - u_y )^2 dxdy}}} },
\]
and the oscillation period compared to the reference solutions.
\begin{table}
\caption{Computed average Nusselt numbers, average velocity norm, average vorticity norm, and oscillation period
         for the tall cavity compared to the reference solutions.
         \label{table:tall-cavity-metrics}
}
\begin{center}
\begin{tabular}{cccD{.}{.}{5}D{.}{.}{4}D{.}{.}{4}D{.}{.}{3}}
$M$          & $N$           & \multicolumn{1}{c}{$\Delta t$} 
                                          & \multicolumn{1}{c}{$\overline{Nu}$} 
                                                           & \multicolumn{1}{c}{$\overline{U}$} 
                                                                               & \multicolumn{1}{c}{$\overline{\omega}$} 
                                                                                               & \multicolumn{1}{c}{$\tau$} \\ \hline
$4\times 20$ & $ 6\times  6$ &   $6.92\times 10^{-3}$   & 4.58356        &  0.2420           &   3.0332      &  3.404    \\
$4\times 20$ & $10\times 10$ &   $6.92\times 10^{-3}$   & 4.57951        &  0.2395           &   3.0172      &  3.411    \\
$4\times 20$ & $14\times 14$ &   $6.92\times 10^{-4}$   & 4.58396        &  0.2421           &   3.0345      &  3.403    \\
$4\times 20$ & $14\times 14$ &   $1.38\times 10^{-3}$   & 4.58393        &  0.2421           &   3.0344      &  3.397    \\
$4\times 20$ & $14\times 14$ &   $2.76\times 10^{-3}$   & 4.58397        &  0.2421           &   3.0342      &  3.403    \\
\multicolumn{3}{c}{Reference solutions:}  & 4.57946        &  0.2397$\dag$     &   2.9998$\dag$&  3.412    \\ \hline
\end{tabular}
\end{center}
$\dag$: The average velocity and vorticity were not given in~%
\cite{SXin_PLeQuere_2002a}
the reference values are the average of 29 solutions presented at the workshop~%
\cite{MAChriston_PMGresho_SBSutton_2002a}\\

\end{table}
\bibliography{references}

\begin{thebibliography}{}

\bibitem[\protect\citename{Christon {\em et~al.},
  }2002]{MAChriston_PMGresho_SBSutton_2002a}
Christon, M.~A., Gresho, P.~M., \& Sutton, S.~B. 2002.
\newblock Computational predictability of time-dependant natural convection
  flows in enclosures (including a benchmark solution).
\newblock {\em Int. J. Numer. Meth. Fluids}, {\bf 40}, 953--980.

\bibitem[\protect\citename{de~Vahl~Davis, }1983]{GdeVahlDavis_1983a}
de~Vahl~Davis, G. 1983.
\newblock Natural convection in a square cavity: {A} bench mark numerical
  solution.
\newblock {\em Int. J. Numer. Meth. Fluids}, {\bf 3}, 249--264.

\bibitem[\protect\citename{de~Vahl~Davis \& Jones,
  }1983]{GdeVahlDavis_IPJones_1983a}
de~Vahl~Davis, G., \& Jones, I.~P. 1983.
\newblock Natural convection in a square cavity: {A} comparison exercise.
\newblock {\em Int. J. Numer. Meth. Fluids}, {\bf 3}, 227--248.

\bibitem[\protect\citename{Fischer {\em et~al.},
  }2000]{PFFischer_NIMiller_FMTufo_2000a}
Fischer, P.~F., Miller, N.~I., \& Tufo, F.~M. 2000.
\newblock An overlapping {Schwarz} method for spectral element simulation of
  three-dimensional incompressible flows.
\newblock {\em In:} Bj{\o}rstad, P., \& Luskin, M. (eds), {\em Parallel
  Solution of Partial Differential Equations}.
\newblock Springer-Verlag.

\bibitem[\protect\citename{Hortmann {\em et~al.},
  }1990]{hortmann-etal:multigrid-benchmark-90}
Hortmann, M., Peri{\'c}, M., \& Scheurer, G. 1990.
\newblock Finite Volume Multigrid Prediction of Laminar Natural Convection:
  Bench-mark Solutions.
\newblock {\em Int. J. Numer. Meth. Fluids}, {\bf 11}, 189--207.

\bibitem[\protect\citename{Maday {\em et~al.},
  }1990]{YMaday_APatera_EMRonquist_1990a}
Maday, Y., Patera, A., \& R{\o}nquist, E.~M. 1990.
\newblock An operator-integration-factor method for time-dependent problems:
  Application to incompressible fluid flow.
\newblock {\em J. Sci. Comput.}, {\bf 4}, 263--292.

\bibitem[\protect\citename{Patera, }1984]{ATPatera_1984a}
Patera, A.~T. 1984.
\newblock A spectral element method for fluid dynamics: Laminar flow in a
  channel expansion.
\newblock {\em J. Comput. Phys.}, {\bf 54}, 468--488.

\bibitem[\protect\citename{Wasberg {\em et~al.},
  }2001]{CEWasberg_OAndreassen_BAPReif_2001a}
Wasberg, C.~E., Andreassen, {\O}., \& Reif, B. A.~P. 2001.
\newblock Numerical simulation of turbulence by spectral element methods.
\newblock {\em Pages  387--402 of:} Skallerud, B., \& Andersson, H.~I. (eds),
  {\em {MekIT}'01. First national conference on Computational Mechanics}.
\newblock Trondheim: Tapir Akademisk Forlag.

\bibitem[\protect\citename{Xin \& {Le~Qu\'{e}r\'{e}},
  }2002]{SXin_PLeQuere_2002a}
Xin, S., \& {Le~Qu\'{e}r\'{e}}, P. 2002.
\newblock An extended {Chebyshev} pseudo-spectral benchmark for the {8:1}
  differentially heated cavity.
\newblock {\em Int. J. Numer. Meth. Fluids}, {\bf 40}, 981--998.

\end{thebibliography}
\end{document}